\documentclass{sf2a-conf2017}
\usepackage{graphicx}
\usepackage{hyperref}
\usepackage[]{natbib}  
\usepackage{amssymb,amsmath,tabularx}

\def\BibTeX{{\rm B\kern-.05em{\sc i\kern-.025em b}\kern-.08em
    T\kern-.1667em\lower.7ex\hbox{E}\kern-.125emX}}
\bibpunct{(}{)}{;}{a}{}{,}  
\newcommand{\kms}{{\mathrm{km~s^{-1}}}}

\newcommand\bb[1] {   \mbox{\boldmath{$#1$}}  }

\begin{document}
\newcolumntype{M}{>{\centering\arraybackslash}m{\dimexpr.5\linewidth-2\tabcolsep}}

\TitreGlobal{SF2A 2017}

\title{Non-adiabatic oscillations of fast-rotating stars: the example of Rasalhague}

\runningtitle{Non-adiabatic oscillations of fast-rotating stars}

\author{G. M. Mirouh$^{1,}$}
\address{SISSA - International School for Advanced Studies, via Bonomea 265, I-34136 Trieste, Italy}
\address{LESIA, Observatoire de Paris, PSL Research University, CNRS, Sorbonne Universit\'es, UPMC Univ. Paris 06, Univ. Paris Diderot, Sorbonne Paris Cit\'e, 5 place Jules Janssen, F-92195 Meudon Cedex, France}
\address{IRAP, Universit\'e de Toulouse, CNRS, UPS, CNES, 14 avenue Edouard Belin, F-31400 Toulouse, France}
\author{D. R. Reese$^{2}$}
\author{M. Rieutord$^{3}$}
\author{J. Ballot$^{3}$}

\setcounter{page}{237}


\maketitle


\begin{abstract}
\end{abstract}


Early-type stars generally tend to be fast rotators. In these stars, mode
identification is very challenging as the effects of rotation are not well
known. We consider here the example of $\alpha$ Ophiuchi, for which dozens of
oscillation frequencies have been measured. We model the star using the
two-dimensional structure code ESTER, and we compute both adiabatic and
non-adiabatic oscillations using the TOP code.  Both calculations yield very
complex spectra, and we used various diagnostic tools to try and identify the
observed pulsations. While we have not reached a satisfactory mode-to-mode
identification, this paper presents promising early results.

\section{Introduction}
In many early-type (O-,B- and A-type) stars, the $\kappa$-mechanism excites both pressure 
and gravity modes. These stars are also usually fast rotators, with an average 
$v\sin i \sim 100-200\kms$ \citep{royer2009}.
The impact of this rotation is twofold: the star is flattened by the centrifugal
force, while the Coriolis acceleration modifies the mode properties. Both of those
effects scramble the oscillation spectra and make mode identification much harder,
requiring two-dimensional stellar models and oscillation calculations.

In this work, we extend the work by \citet{mirouh_etal14a} by using the so-called 
forward approach to model the fast rotator Rasalhague ($\alpha$ Ophiuchi).
We model the star using the two-dimensional code ESTER 
that fully takes rotation into account, but leaves out mass loss and chemical diffusion
in the star, and then compute the non-perturbative oscillation spectrum  
in the same geometry using the TOP program. Due to the high resolution used to describe
the stellar interior and the modes, this typically yields several hundreds of modes, 
among which we need to select the most relevant candidates for identification.
Another approach consists in looking for regular patterns in the oscillation
spectrum: because of the effects of rotation, theoretically-predicted patterns \citep{LG09} 
went long unnoticed, until \citet{AGH2015} was able to find a large separation
in a small sample of stars and link it to stellar properties.

\section{Rasalhague}
Our study case is the A-type star Rasalhague.  
Interferometry showed that the star
is seen almost equator-on, with a ratio between the equatorial and polar radii of $1.2$,
i.e. $R({\rm pole})=1.2R({\rm equator})$ \citep{zhao_etal09}.  
Its mass has been constrained by means of its orbiting companion
\citep{hinkley_etal11} and interferometry to $M \sim 2.18 - 2.4 M_\odot$.
Moreover, fifty-seven oscillation frequencies, ranging from $1.768$ to $48.347$ c/d, 
have been measured with the asteroseismology mission MOST \citep{monnier_etal10}.

We compute an ESTER model \citep{ELR13, RELP2016} that fits the luminosity and the 
equatorial and polar radii, the properties of which are summarized in table \ref{tab:model}.
This model is the same as the one presented in \citet{mirouh_etal14a}.

\begin{table}[h]
$$
\begin{array}{|c|c||c|c|}
\hline
\Omega / \Omega_K               & 0.624         & T_{\rm eff}(p) (K)              & 9177\\
M / M_\odot                     & 2.22          & T_{\rm eff}(e) (K)              & 7731\\
X_c/X                           & 0.3685        & R(p)/ R_\odot                   & 2.388\\
X                               & 0.7           & R(e)/ R_\odot                   & 2.858\\
Z                               & 0.02          & V_{\rm eq} ({\rm km.s}^{-1})    & 240  \\
                                &               & L/L_\odot                       & 31.1 \\
\hline
\end{array}
$$
\caption{Properties of our model for Rasalhague ($\alpha$ Ophiuchi). 
  The left part of the table contains the input parameters of the model, while
  the right part lists the quantities matched to the quantities observed for the star.
  }
\label{tab:model}
\end{table}

\section{Adiabatic oscillations}
\begin{figure}[h!]
  \includegraphics[width=\textwidth]{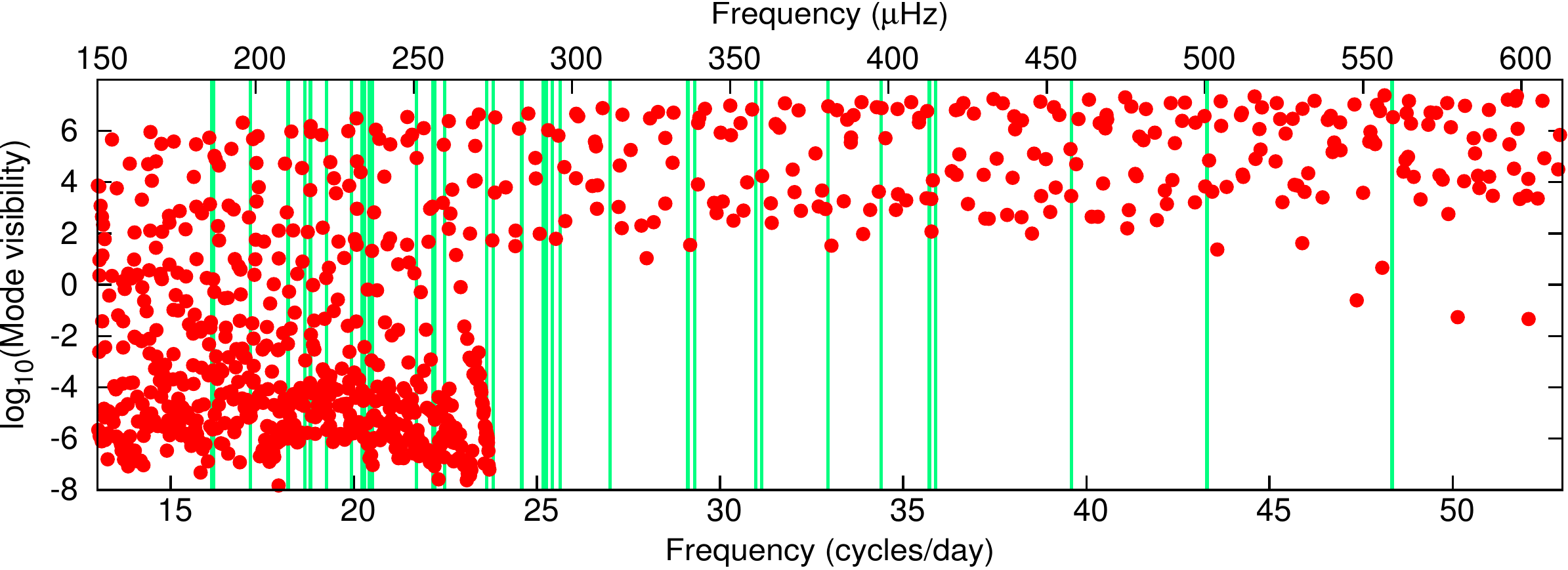} \hfill 
  \caption{Mode visibilities of $m=0$ modes, obtained through adiabatic calculations. 
    The low-frequency low-visibility clump corresponds mostly to g
    modes, while the higher-visibility strip spanning the whole frequency range
    corresponds to p modes. Each vertical green line corresponds to a frequency
    measured by \citet{monnier_etal10}.}
  \label{fig:visib}
\end{figure}

The eigenvalue problem of adiabatic oscillations is solved with
the TOP code \citep{RTMJSM09} for modes with azimuthal orders $-4\leq m\leq 4$, in the
range of frequencies in which modes are observed. 
We find g modes and p modes modified by rotation, and we can clearly
distinguish these two populations in the spectra (see Figure \ref{fig:visib}).

To select the modes that might be seen from Earth, we compute the mode visibilities,
following \citet{reese_etal13}. As the mode amplitudes cannot be derived from a 
linear calculation, we need to normalize the eigenmodes. For this purpose, we normalize
all our solutions by $\max||\bb{\xi}||^2 \left( \omega + m\Omega\right)^2$, where $\bb{\xi}$ is the Lagrangian
displacement and $\left(\omega + m\Omega\right)$ the corotating mode frequency \citep{reese_etal17}.
We find that g modes are the least visible, as one would expect, considering
they probe deep layers of the star and are usually evanescent towards the surface. 
Located near the surface, p modes are much more visible. 

We also compute the thermal dissipation rates using the quasi-adiabatic
approximation \citep{unno_etal89}, which yields only linearly stable modes. 
Using this method, the more visible p modes appear to be more damped than the 
g modes. This conundrum seems to prevent the identification of the modes.

\section{Non-adiabatic oscillations}
\begin{figure}[h!]
  \includegraphics[angle=-90,width=\textwidth]{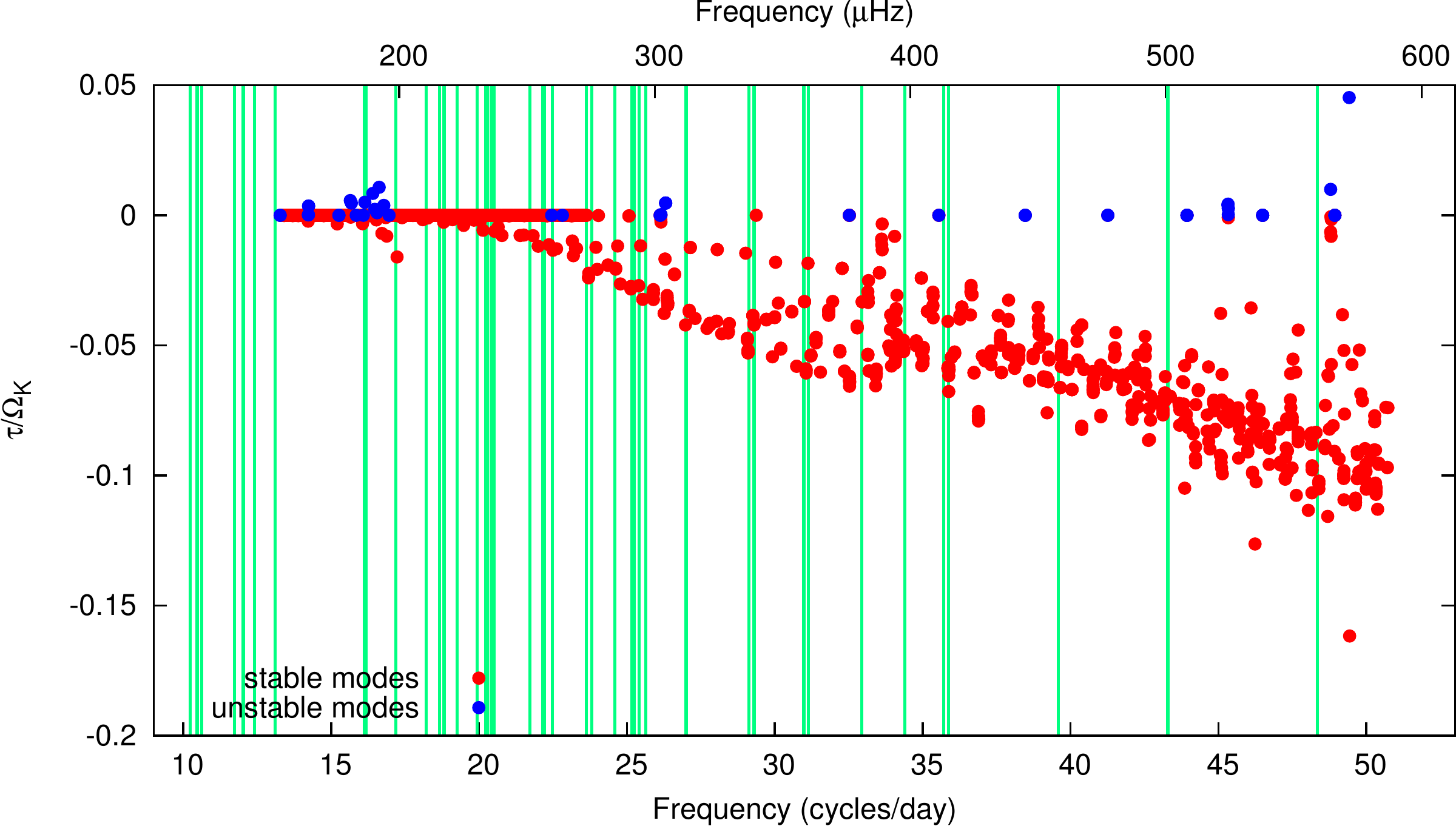} \hfill 
  \caption{Damping rates for the axisymmetric modes ($m=0$) obtained through
  the non-adiabatic calculation.  We find amplified g and p modes, most of the
  latter seem to follow a regular pattern.  }
  \label{fig:damp}
\end{figure}

In an attempt to improve our description of the modes, we use the non-adiabatic 
version of TOP \citep{reese_etal16}.
This calculation is much more demanding numerically but results in complex eigenvalues 
that include both the modes' frequencies and growth rates.
While the mode visibilities present the same properties as that of the adiabatic modes
shown in figure \ref{fig:visib}, we are able to find amplified modes. 
Figure \ref{fig:damp} shows the growth rates of the modes.
Keeping only the most visible amplified modes does not seem to allow a direct
identification of the modes, but fine tuning the models might lead to a
better match. 
The ESTER models also suffer from a couple of limitations that may affect
our calculation: as of now, the code does not implement evolution but determines the stellar
structure at a given time; it also leaves aside any surface convection. 
As this prevents us from describing accurately the
transfer of heavy elements from the core to the envelope due to core recession
or diffusive effects along the main-sequence evolution, and results in a two-domain 
chemical composition with a depleted core and a homogeneous envelope, it may 
impact individual eigenmodes, and especially the g modes that probe deep layers in the star.

We also find a series of amplified modes that seem spaced by a regular separation.
However, this separation is not exactly constant ($\sim 30 \mu$Hz) and does not
seem to match the one uncovered by \citet{AGH2015} ($20\pm 1 \mu$Hz).
When checking the nature of the modes, it appears that these amplified modes are
high-degree gravity modes, and not the island pressure modes hypothesized by 
\citet{LG09}.

Two-dimensional non-adiabatic calculations are still at an early stage, and the
needed resolution, associated with the high number of variables and equations
involved, impact the precision of the solutions. To improve the numerical
convergence, we decide to split the star in two domains: we solve the perturbed
equations using the adiabatic approximation in the inner part of the star,
while the full non-adiabatic equations are solved in the outer part.  
For a first exploration, we set the Lagrangian perturbation of entropy to zero
at the interface between the two domains. Figure 3 shows the 
dispersion of the solutions for various sizes of the inner adiabatic domain,
for the p mode at $f= 424.378 \mu$Hz when using slightly different input guesses. 

We see that as we move the interface outwards, the solutions first get closer
to the average value (that we suppose to be correct): this is explained by the
disappearance of numerical errors, as the very small $\delta s$ is no longer
computed in the inner part of the star but set to zero by hand.  However, when
the adiabatic region becomes larger, the growth rate gradually falls to zero,
as we start neglecting non-adiabatic effects that are important in outer
layers.

\begin{figure}[h!]
  \begin{tabular}{MM}
    \includegraphics[angle=-90,width=0.55\textwidth]{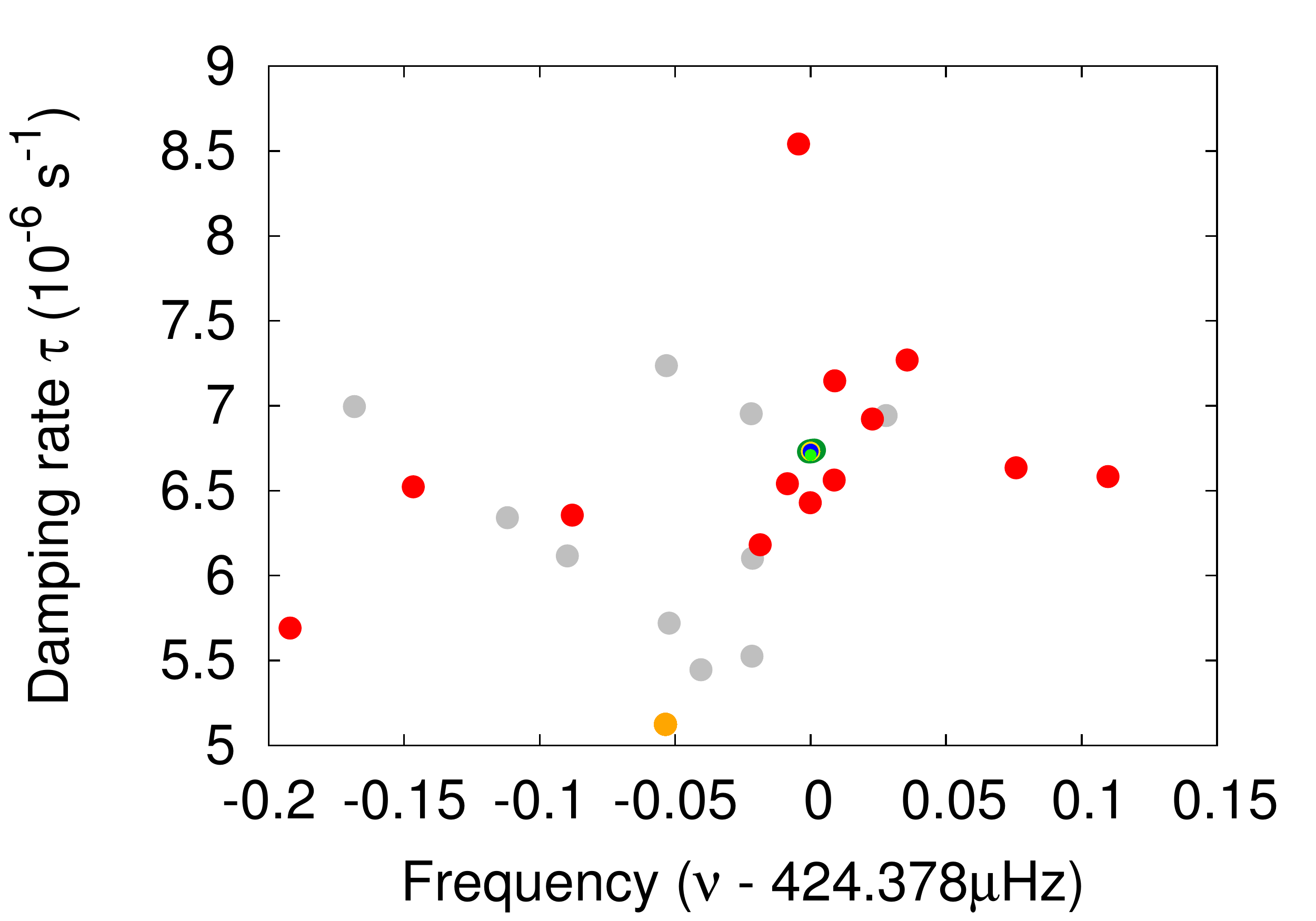} &
    \includegraphics[width=0.35\textwidth]{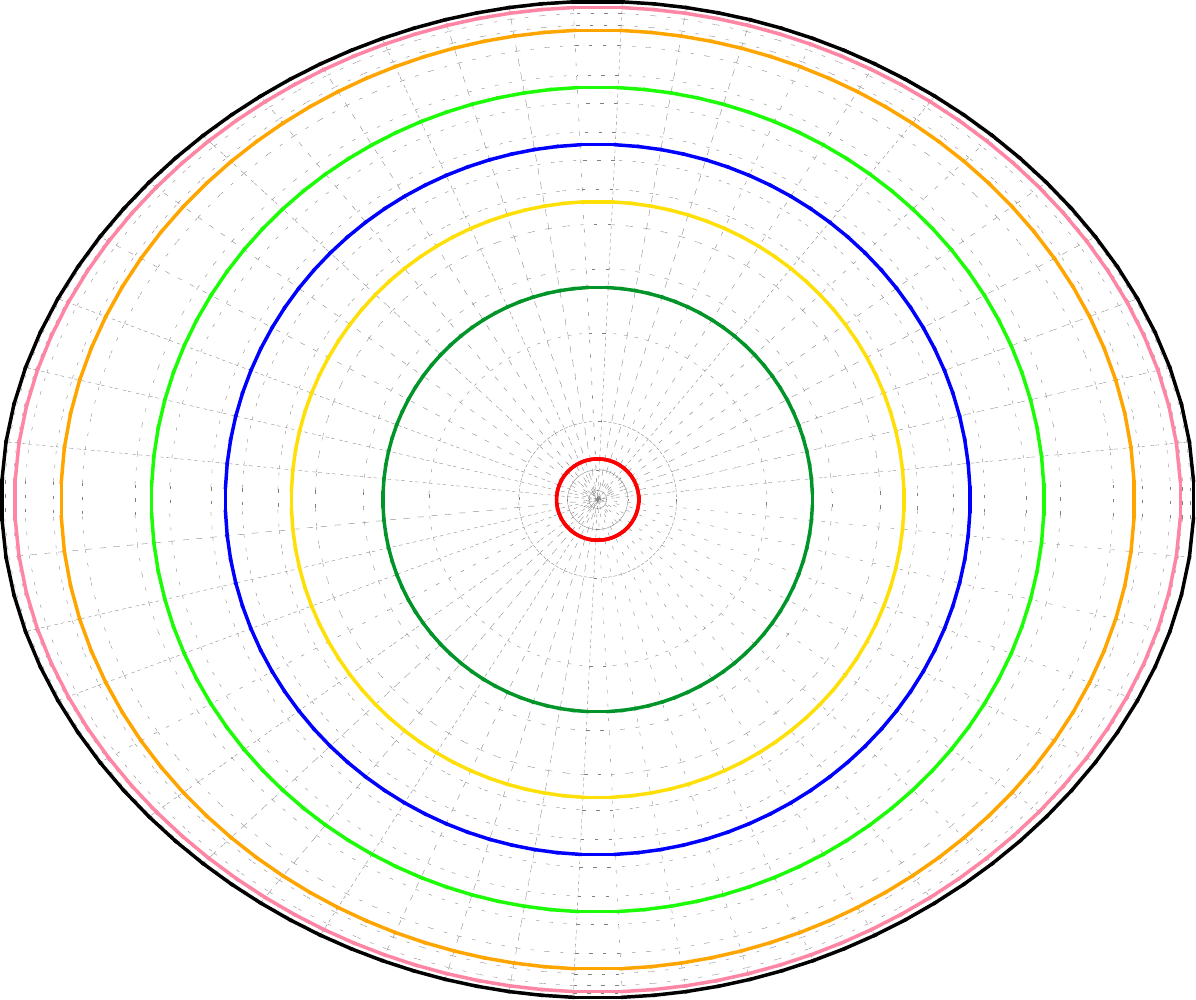}
  \end{tabular}
  \caption{The left plot shows the dispersion for the split calculation. 
    We place the adiabatic inner domain/non-adiabatic outer domain interface
    at various depths, shown on the meridional cut on the right plot.  The
    colours of the symbols correspond to those of the interfaces:
    the grey points corresponding to the full non-adiabatic calculation, while
    the dark green, yellow, blue and green dots overlap. We
    omit the result of the fully adiabatic calculation, which would have 
    $\tau = 0$.}
  \label{fig:disp}
\end{figure}

\section{Conclusions and future prospects}
In this work, we computed a two-dimensional model and both the adiabatic and non-adiabatic
oscillations of the fast-rotating $\delta$ Scuti star $\alpha$ Ophiuchi, in order to identify
the observed oscillation frequencies.
Because of the complexity of the oscillation spectrum, and the high number of calculated 
eigenmodes, we computed the damping rates and mode visibilities in order to narrow down 
the range of possible matches between calculated and observed modes.
The adiabatic calculation, using the so-called quasi-adiabatic approximation predicts 
only damped modes. This is due to a poor description of the modes near the stellar surface.
Using a non-adiabatic calculation allows us to find amplified modes. As we cannot predict 
the amplitude of the modes, we normalize them and compute their visibilities. 
Using these two diagnostics, we are able to select only a reasonably small subsample of the
whole synthetic spectrum, that we compare to the observations.
At this point, we were unable to identify individual frequencies or find
regular patterns that match the observations.
This may come from the numerical issues raised by the non-adiabatic
calculation or the limitations of the ESTER stellar models. Improving both of these aspects will
be necessary to make the oscillation growth rates and the rotating star structures more
reliable and reach a satisfying two-dimensional seismic inference for Rasalhague.


\bibliography{mirouh}
\bibliographystyle{aa}  
%
%
%
\end{document}